\documentclass[sigconf]{acmart}
\AtBeginDocument{%
  }

\usepackage[capitalise]{cleveref}
\creflabelformat{equation}{#2#1#3}

\usepackage{tabularx}

\copyrightyear{2025}
\acmYear{2025}
\setcopyright{acmlicensed}\acmConference[GECCO '25]{Genetic and
Evolutionary Computation Conference}{July 14--18, 2025}{Malaga, Spain}
\acmBooktitle{Genetic and Evolutionary Computation Conference (GECCO '25),
July 14--18, 2025, Malaga, Spain}
\acmDOI{10.1145/3712256.3726402}
\acmISBN{979-8-4007-1465-8/2025/07}

\begin{document}

%%
%% The "title" command has an optional parameter,
%% allowing the author to define a "short title" to be used in page headers.
\title{Evaluating Mutation Techniques in Genetic Algorithm-Based Quantum Circuit Synthesis}

%%
%% The "author" command and its associated commands are used to define
%% the authors and their affiliations.
%% Of note is the shared affiliation of the first two authors, and the
%% "authornote" and "authornotemark" commands
%% used to denote shared contribution to the research.
\author{Michael Kölle}
\affiliation{%
  \institution{LMU Munich}
  \city{Munich}
  \country{Germany}}
\email{michael.koelle@ifi.lmu.de}

\author{Tom Bintener}
\affiliation{%
  \institution{LMU Munich}
  \city{Munich}
  \country{Germany}}

\author{Maximilian Zorn}
\affiliation{%
  \institution{LMU Munich}
  \city{Munich}
  \country{Germany}}
  
\author{Gerhard Stenzel}
\affiliation{%
  \institution{LMU Munich}
  \city{Munich}
  \country{Germany}}
  
\author{Leo Sünkel}
\affiliation{%
  \institution{LMU Munich}
  \city{Munich}
  \country{Germany}}

\author{Thomas Gabor}
\affiliation{%
  \institution{LMU Munich}
  \city{Munich}
  \country{Germany}}

\author{Claudia Linnhoff-Popien}
\affiliation{%
  \institution{LMU Munich}
  \city{Munich}
  \country{Germany}}

%%
%% By default, the full list of authors will be used in the page
%% headers. Often, this list is too long, and will overlap
%% other information printed in the page headers. This command allows
%% the author to define a more concise list
%% of authors' names for this purpose.
\renewcommand{\shortauthors}{Kölle et al.}

%%
%% The abstract is a short summary of the work to be presented in the
%% article.
\begin{abstract}
Quantum computing leverages the unique properties of qubits and quantum parallelism to solve problems intractable for classical systems, offering unparalleled computational potential. However, the optimization of quantum circuits remains critical, especially for noisy intermediate-scale quantum (NISQ) devices with limited qubits and high error rates. Genetic algorithms (GAs) provide a promising approach for efficient quantum circuit synthesis by automating optimization tasks.
This work examines the impact of various mutation strategies within a GA framework for quantum circuit synthesis. By analyzing how different mutations transform circuits, it identifies strategies that enhance efficiency and performance. Experiments utilized a fitness function emphasizing fidelity, while accounting for circuit depth and T operations, to optimize circuits with four to six qubits. Comprehensive hyperparameter testing revealed that combining delete and swap strategies outperformed other approaches, demonstrating their effectiveness in developing robust GA-based quantum circuit optimizers.
\end{abstract}

%%
%% The code below is generated by the tool at http://dl.acm.org/ccs.cfm.
%% Please copy and paste the code instead of the example below.
%%


%%
%% Keywords. The author(s) should pick words that accurately describe
%% the work being presented. Separate the keywords with commas.
%% A "teaser" image appears between the author and affiliation
%% information and the body of the document, and typically spans the
%% page.
% \begin{teaserfigure}
%   \includegraphics[width=\linewidth]{sampleteaser}
%   \caption{Seattle Mariners at Spring Training, 2010.}
%   \Description{Enjoying the baseball game from the third-base
%   seats. Ichiro Suzuki preparing to bat.}
%   \label{fig:teaser}
% \end{teaserfigure}

\received{20 January 2025}
%\received[revised]{12 March 2009}
%\received[accepted]{5 June 2009}

%%
%% This command processes the author and affiliation and title
%% information and builds the first part of the formatted document.
\maketitle

\section{Introduction}\label{sec:introduction}
Quantum computing offers a promising way to tackle complex problems that classical computers cannot solve \cite{shor1999polynomial,kitaev2002classical,grover1996fast}. Quantum circuits lie at the core of quantum computing and are essential for implementing quantum algorithms \cite{chiribella2008quantum}. The efficient synthesis and optimization of these circuits is critical yet challenging, especially on NISQ devices, which have limited qubits and high error rates \cite{pelofske2022quantum,preskill2018quantum}. As more performant quantum hardware continues to emerge, efficient and automated quantum circuit synthesis grows increasingly important, because manual circuit creation is not sustainable \cite{wu2020qgo,biamonte2017quantum,benedetti2019parameterized}.

The synthesis of quantum circuits is difficult because of the complexity of quantum operations and the limits of current hardware. GAs offer a possible solution by using evolutionary strategies to improve quantum circuit designs \cite{shende2005synthesis}. GAs are especially effective at exploring large, complex solution spaces \cite{mathew2012genetic,coello2007evolutionary}, but their performance in quantum circuit synthesis and optimization remains an open area of research \cite{ruican2008genetic,miranda2021synthesis}. In particular, examining how different mutation strategies interact with quantum circuits may provide new ways to enhance optimization efficiency.

To investigate these mutation strategies, we conducted a series of experiments. We developed a multifunctional quantum environment that can create, manipulate, and evaluate quantum circuits efficiently. It includes a circuit optimizer that works within the Clifford + T set, following standard protocols \cite{selinger2012efficient}, and an unbiased circuit generator that produces diverse candidate circuits for testing. An autonomous system switches between serial and parallel processing modes based on available hardware and data, ensuring optimal resource use.

A GA framework was then implemented within this quantum environment. The fitness function in this framework is derived from fidelity, circuit depth, and the number of T operations \cite{liang2019quantum,zhang2022quantum,steane1999efficient}. The framework allows either a single-population or an island model, along with tournament selection for elites and offsprings, immigrants, and single-point crossover \cite{whitley1999island,miller1995genetic,kora2017crossover}. The mutation strategies include change, delete, add, swap, and every possible combination of these, using either static or dynamic mutation rates. Hyperparameters are tuned autonomously after each repetition to improve performance. Six-qubit circuits form the dataset for these evaluations, and the same dataset is reused to maintain consistency in performance measurements.

Our contributions involve a comprehensive, modular quantum environment that serves as a framework for evaluating GA-based and other machine learning approaches in quantum computing. We also supply empirical data on how various mutation strategies perform in quantum synthesis and offer guidance on their efficient application.

The remainder of this work is structured as follows: In \cref{sec:related_work}, we review related work on quantum state preparation, quantum circuit optimization, and the application of genetic algorithms in quantum computing. \cref{sec:env} introduces our quantum circuit environment. \cref{sec:exp-setup} describes the experimental setup. \cref{sec:results} presents the results, analyzing the performance of the genetic algorithm, the impact of mutation strategies, and the effects of mutation rate, population size, and adaptive mutation. Finally, \cref{sec:conclusion} concludes the paper, summarizing key findings and discussing potential directions for future research.

\section{Related Work}\label{sec:related_work}
This section provides an overview of the foundational and recent research that underpins this work. It focuses on quantum state preparation, quantum circuit optimization, quantum circuit synthesis, and the application of GAs in quantum computing. 

\subsection{Quantum State Preparation}\label{sec:quantum_state_preparation}
Quantum hardware typically initializes each qubit in the \(|0\rangle\) state by default, yet quantum algorithms often require more complex states. Preparing these states manually is time-intensive, especially for multi-qubit algorithms, and does not guarantee an optimal solution. To automate this process, researchers have proposed several methods, including state decomposition \cite{mottonen12006decompositions}, black-box approaches \cite{sanders2019black}, and adiabatic methods \cite{coello2022quantum}. 

Another prominent direction uses machine learning algorithms to train parameterized circuits that generate specific target states. Recent papers demonstrate that reinforcement learning agents can efficiently assemble state preparation sequences \cite{kolle2024reinforcement,porotti2022deep}, reinforcing the notion that goal-oriented methods are well-suited for automating state preparation.

\subsection{Quantum Circuit Optimization and Synthesis}\label{sec:quantum_circuit_opt_and_syn}
Quantum circuit optimization and synthesis are core techniques in state preparation. Optimization aims to reduce the number of operations in existing circuits by removing redundancies, reordering gates, or minimizing T-gates \cite{ge2024quantum}. This step is crucial in NISQ-era hardware, where qubit counts are low and noise is high. Reducing circuit depth and gate count mitigates noise, improving overall reliability.

Synthesis centers on decomposing target states into smaller subcircuits or template-based solutions. Efficient decomposition and template matching help researchers and automated tools build complex circuits more readily \cite{ge2024quantum}. Despite ongoing progress, the search for better optimization and synthesis strategies continues, including contributions from the present work.

\subsection{Genetic Algorithms in Quantum Computing}\label{sec:genetic_algorithms_in_quantum_computing}
Inspired by the successful use of reinforcement learning in state preparation, researchers have also explored GAs for optimizing quantum circuits. GAs excel at navigating large search spaces, making them promising candidates for quantum state preparation tasks \cite{creevey2023gasp,rindell2023exploring,wright2024t}. 
Miranda et al. \cite{miranda2021synthesis} notably demonstrated the effectiveness of an island-model GA, in which the population splits into partially isolated subpopulations. This division helps avoid premature convergence by maintaining diversity across subgroups. Sunkel \cite{sunkel2023ga4qco} introduced a comprehensive GA framework that offers standardized interfaces and extensibility, enabling future enhancements without altering core components. 
Meanwhile, Ge \cite{ge2024quantum} presented guidelines for crafting fitness functions and prioritizing circuit properties in the NISQ era, where limited qubit counts and elevated error rates constrain feasible circuit sizes. This work also highlighted metrics for mitigating noise and error accumulation, and provided strategies for balancing gate fidelity against circuit depth to ensure viability on near-term hardware.

\section{Quantum Circuit Environment}\label{sec:env}
\begin{figure}[htb]
    \centering
    \includegraphics[width=\linewidth]{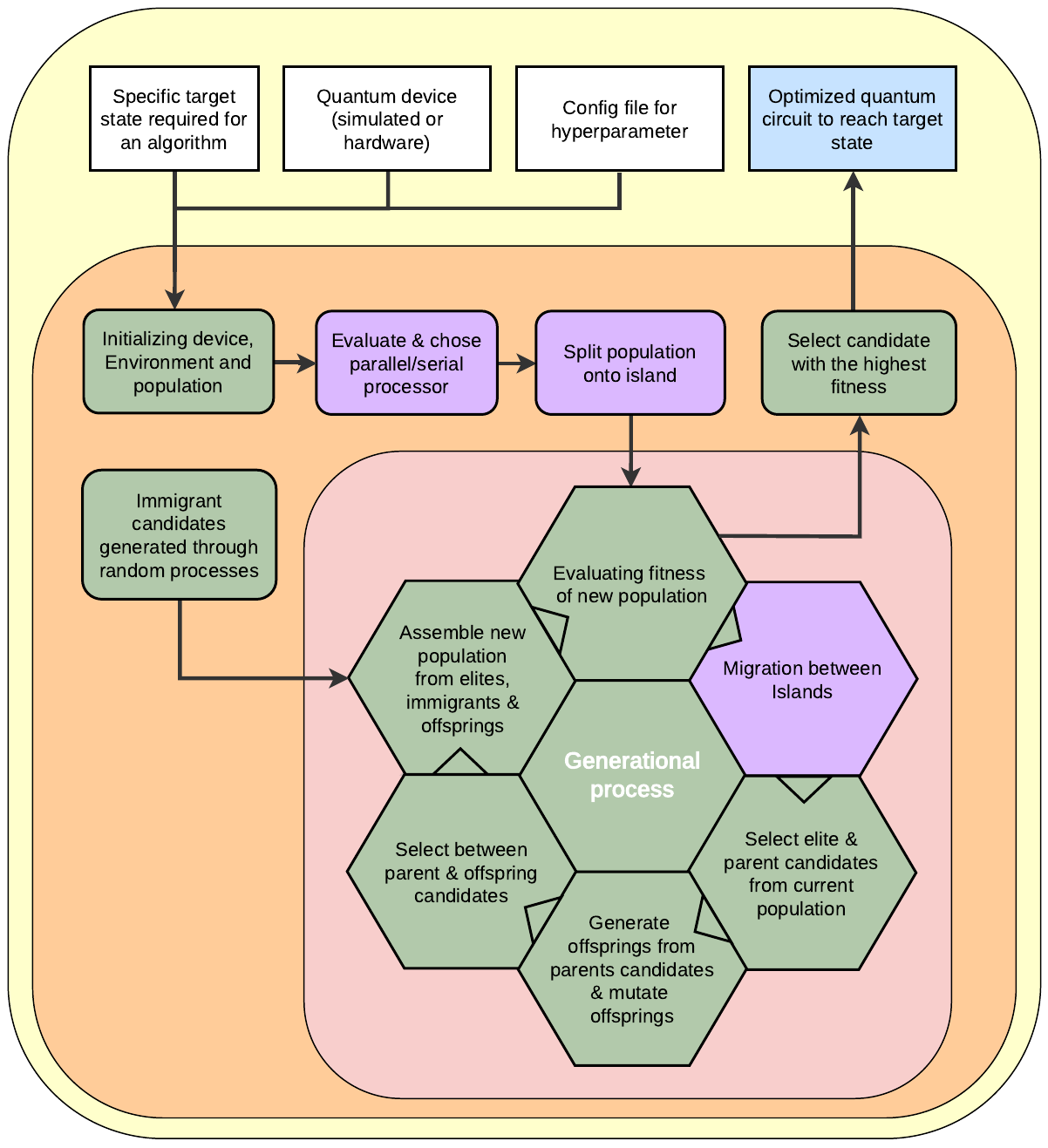}
    \caption{The GA optimization process. Yellow represents the outer layer, orange represents GA steps, and red represents the repeated optimization cycle. White boxes indicate inputs, blue boxes indicate outputs, green boxes indicate essential steps, and purple boxes indicate optional steps.}
    \Description{The GA optimization process}
    \label{fig:ga_optimization_process}
\end{figure}

This section introduces the quantum circuit environment developed for synthesizing quantum states with a GA. It describes the GA structure and outlines how optimization proceeds for a given problem. The goal is to explain the main components and their implementation. This environment uses flexible approaches to candidate generation, mutation, and evaluation to adapt to diverse quantum state synthesis tasks. \cref{fig:ga_optimization_process} presents the streamlined optimization process used in this work, highlighting the steps for refining candidate circuits toward the target quantum state.

\subsection{Candidate Representation}\label{sec:candidate_representation}
\begin{figure}[htb]
    \centering
    \includegraphics[width=0.8\linewidth]{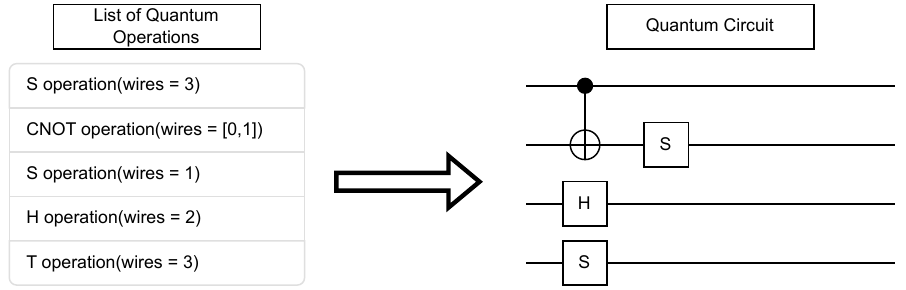}
    \caption{An example list of quantum operations and its corresponding circuit.}
    \Description{An example list of quantum operations and its corresponding circuit}
    \label{fig:list_to_circuit}
\end{figure}

A candidate \(C\) in a population \(P\) is a potential solution for an optimization problem \(O\). Each \(C_i\) with \(0 < i < |P|\) is stored as a one-dimensional list of quantum operations:
\begin{itemize}
    \item \textbf{Id:} A label denoting a quantum operation (e.g., Hadamard, CNOT).
    \item \textbf{Wires:} A number or list specifying which qubit(s) the operation targets.
\end{itemize}
\cref{fig:list_to_circuit} shows how a list of operations translates into a quantum circuit. This list-based format simplifies iteration, manipulation, and assessment when performing genetic operations such as mutation and crossover.

\subsection{Target State}\label{sec:target_state}
The target state is the desired quantum state for a specific algorithm or experiment, and it defines the objective in quantum circuit optimization. Each candidate’s effectiveness is measured by its success in producing this state. A target state can be characterized by its statevector and density matrix. %, as in \cref{state_vector} and \cref{densitymatrix}, respectively.

\subsection{Environment Initialization}\label{sec:environment_initialization}
The genetic algorithm requires three key inputs to begin: a target density matrix, which specifies the desired quantum state and thus defines the optimization goal; a quantum device (either simulated or physical) to execute and evaluate the circuits; and a configuration file containing GA parameters such as generation count, stopping criteria, population size, and mutation rates. To simplify customization, these parameters are organized into distinct sections, including run, population, island, fitness, evolutionary, and parallel. For instance, the population section defines the number of candidates and the minimum and maximum circuit depths, while the island section governs the number of islands and the frequency of candidate migration between them. This structured approach makes it easier to adjust the GA for different experimental setups and performance objectives.

\subsection{Population Initialization}\label{sec:population_initalisation}
\begin{figure}[htb]
    \centering
    \includegraphics[width=0.85\linewidth]{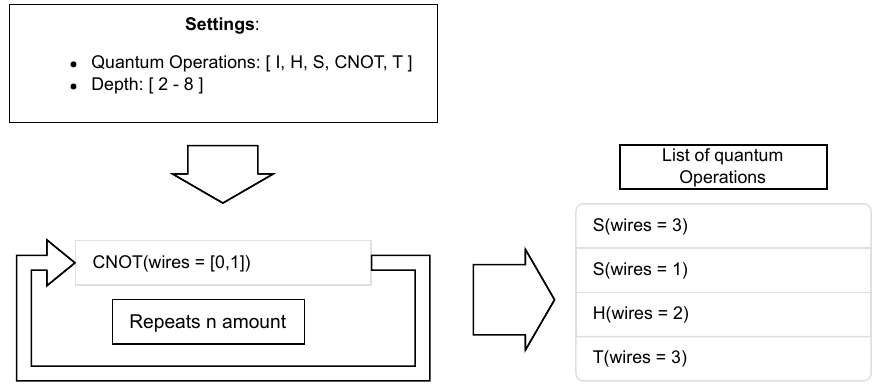}
    \caption{Using Python lists to define circuit depth and operations for each candidate during population initialization.}
    \Description{Using Python lists to define circuit depth and operations for each candidate during population initialization}
    \label{fig:operations_to_list}
\end{figure}

Population initialization produces an initial set of candidates on which future generations build. \cref{fig:operations_to_list} shows the process of generating random quantum circuits by choosing circuit depth, quantum operations, and qubit assignments. A short post-processing step removes trivial candidates lacking sufficient complexity. This approach provides a diverse initial population and lays the groundwork for effective optimization.

\subsection{Evaluation Process}\label{sec:evaluation_processor}
To evaluate candidates, the framework offers three main methods that target different circuit configurations: Parallel processing, which runs batches of candidates concurrently on multicore processors, is well-suited to circuits with four or more qubits. In contrast, serial single processing evaluates candidates individually and minimizes overhead, making it optimal for circuits with fewer than four qubits and shallow depth. For smaller circuits that have higher depth, serial batch processing assesses all candidates in a single batch, reducing redundancy. If the configuration does not specify a particular method, the framework runs a quick test on the initial population to determine which approach completes fastest and then applies that method for all remaining generations.

\subsection{Fitness Evaluation}\label{sec:fitness_evaluation}
The fitness function incorporates fidelity, circuit depth, and the number of T operations to quantify how well a circuit meets hardware constraints while approximating the target state. The fidelity score, which measures how closely a candidate approximates the target state, is the primary metric for an optimal solution.
\begin{equation}
F(\rho, \sigma) = \left( \operatorname{tr} \left( \sqrt{\sqrt{\rho} \sigma \sqrt{\rho}} \right) \right)^2
\label{fidelity_equation}
\end{equation}
where \(\rho\) and  \(\sigma\) are both density matrices. For practical use, a score between 0.90-0.99 is typically aimed for. Circuit depth measures sequential gate execution; shallower circuits reduce noise accumulation. The number of T gates is another key factor, as T gates are resource-intensive in the Clifford+T gate set. The overall fitness is computed as:
\begin{equation}
s_\text{fitness} = w_{\text{fidelity}} \cdot s_{\text{fidelity}} \;-\; w_{\text{d}} \cdot s_{\text{d}} \;-\; w_{\text{T-ops}} \cdot s_{\text{T-ops}},
\label{fitness_equation}
\end{equation}
where \(w\) is the weight for each metric, \(d\) the circuit depth and \(s\) is the corresponding score.

\subsection{Evolutionary Step}\label{sec:evolutionary_step}
During each generation, the GA refines the population to improve solutions. Offspring form the bulk of the new population and are created through crossover, where randomly selected parents are split, recombined, and mutated. This process introduces new genetic material, promoting diversity while preserving successful traits. Meanwhile, elites—top-performing candidates—advance directly to the next generation without changes, ensuring that high-fidelity solutions remain intact. To maintain diversity and prevent early convergence, a subset of immigrants is randomly generated and included in the population, offering new genetic variations that might otherwise not emerge.

\begin{figure}[htb]
    \centering
    \includegraphics[width=\linewidth]{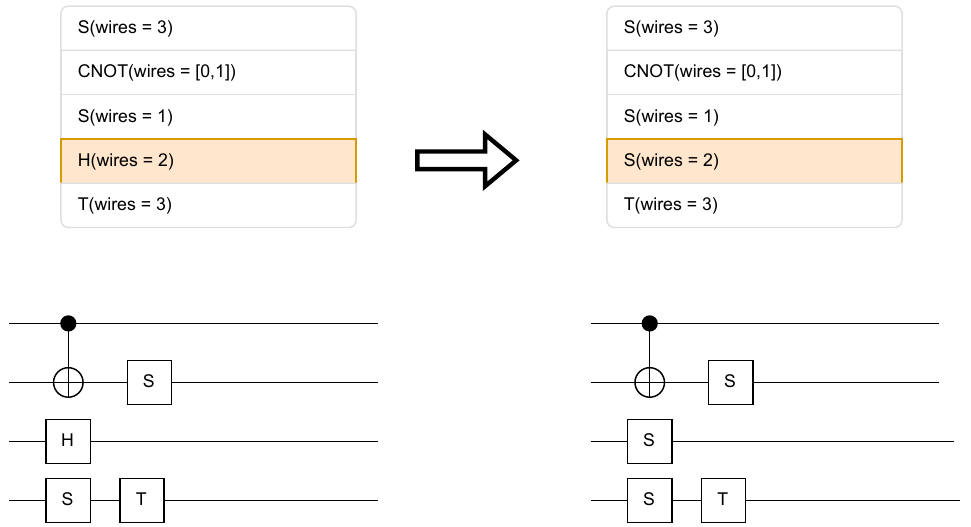}
    \caption{An example of the \emph{change} strategy. The modified operation is highlighted in orange.}
    \Description{An example of the change strategy}
    \label{fig:change}
\end{figure}

\begin{figure}[htb]
    \centering
    \includegraphics[width=\linewidth]{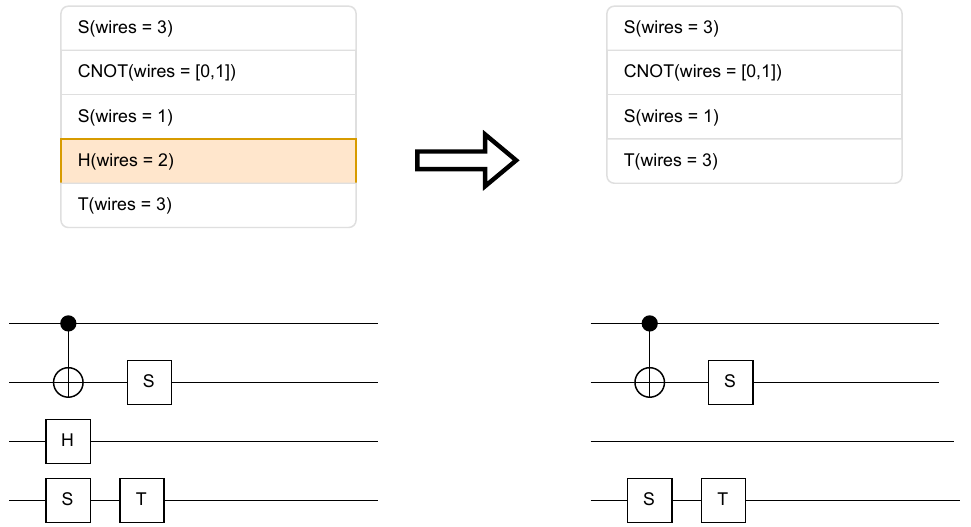}
    \caption{An example of the \emph{delete} strategy. The removed operation is highlighted in orange.}
    \Description{An example of the delete strategy}
    \label{fig:delete}
\end{figure}

\begin{figure}[htb]
    \centering
    \includegraphics[width=\linewidth]{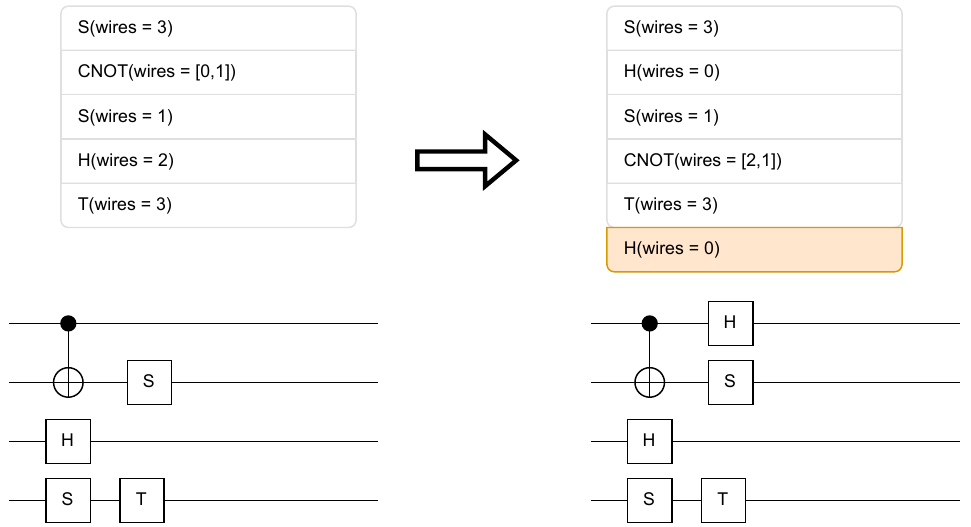}
    \caption{An example of the \emph{add} strategy. The newly inserted operation is highlighted in orange.}
    \Description{An example of the add strategy}
    \label{fig:add}
\end{figure}

\begin{figure}[htb]
    \centering
    \includegraphics[width=\linewidth]{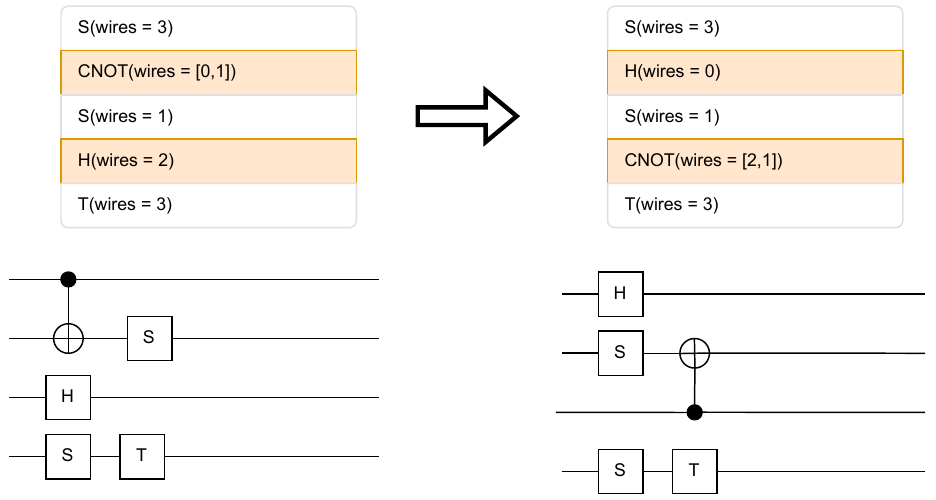}
    \caption{An example of the \emph{swap} strategy. The exchanged operations are highlighted in orange.}
    \Description{An example of the swap strategy}
    \label{fig:swap}
\end{figure}

Mutations then refine candidates by applying minimal changes to their structure. In the \emph{change} mutation, one quantum operation is substituted with another of the same length, ensuring that the circuit depth remains constant (\cref{fig:change}). The \emph{delete} mutation removes an operation to simplify the circuit (\cref{fig:delete}), while the \emph{add} mutation introduces a new operation at a random position, increasing circuit complexity (\cref{fig:add}). Finally, the \emph{swap} mutation exchanges two operations without altering their type (\cref{fig:swap}). This combination of offspring creation, elite preservation, immigrant introduction, and incremental mutations enables the GA to explore the solution space thoroughly while retaining strong candidates for subsequent iterations.


An experimental feature is adaptive mutation, which adjusts mutation parameters in response to average fitness, population diversity, and remaining generations. It encourages extensive exploration early on, then converges toward stronger solutions as the optimization nears completion.

\subsection{Island Model}\label{sec:island_model}
The island model is an optional feature that partitions the population into smaller, parallel subpopulations. Each island explores different regions of the solution space in isolation. During later generations, selective migration shares high-fidelity candidates between islands, boosting convergence toward globally optimal solutions. This parallel isolation also mitigates premature convergence by maintaining diverse subpopulations.

\subsection{Quantum Circuit Optimization}\label{sec:circuit_optimization}
Fully random candidate creation can produce extraneous or canceling operations. To address this, the environment includes a selective optimizer. It merges or removes redundant gates in each candidate’s circuit (e.g., converting two consecutive T gates to an S gate) during fitness evaluation. The best candidate also undergoes a final optimization step before deployment. This selective application retains GA-driven improvements without incurring excessive computational overhead for every generation.

\subsection{Tools and Libraries}\label{sec:tools_and_liberaries}
This environment relies on several well-established libraries to maintain reliability and comparability. PennyLane, an open-source Python framework, enables seamless quantum programming on both simulated and real hardware, allowing the easy translation of operation lists into executable circuits. NumPy supports efficient handling of matrices and arrays, reducing generation times and enabling large-scale simulations. Optuna automates hyperparameter tuning for the GA, storing experiment results in a manageable database that facilitates performance analysis. Lastly, Slurm distributes computational jobs across multiple server nodes, enabling parallel processing and accelerating large-scale or repeated optimizations. Taken together, these tools create a robust and adaptable environment for generating, evaluating, and improving candidate circuits across various mutation settings.

\section{Experimental Setup}\label{sec:exp-setup}
This section describes the decisions made to set up the GA evaluations for quantum circuit optimization. It covers how the evaluations were conducted, including computational resources, dataset generation, hyperparameter tuning, and performance metrics. Each element aims to create a robust, reproducible framework that can assess the GA’s effectiveness across diverse scenarios.

\subsection{Dataset}\label{sec:dataset}
The datasets used in this work comprise randomly generated PennyLane circuits, subsequently optimized as noted in \cref{sec:circuit_optimization}. Each dataset contains circuits with a fixed number of qubits yet varying circuit depths to resemble realistic quantum states. The qubit count ranges from five to eight, as fewer than five qubits often resulted in near-optimal solutions that showed little variation among mutation methods, and more than eight qubits demanded prohibitive computational resources. Circuit depths span from five to fifteen operations, providing enough complexity to challenge the GA while remaining computationally tractable.

\subsection{Performance Metric}\label{sec:performance_metric}
The performance metric measures the GA’s effectiveness in optimizing quantum circuits. Specifically, it is defined as the average of the highest final fitness scores across all circuits in a dataset. This metric evaluates both solution quality and consistency under varying datasets and optimization conditions.

\subsection{Hyperparameter Optimization}\label{sec:hyperparameter_optimization}
Hyperparameter optimization proceeded in two stages. The first stage focused on initialization settings, such as average, minimum, and maximum circuit depths, and tested different numbers of islands for a generic target circuit with four to eight qubits. Wide parameter bounds in Optuna enabled a broad search to establish a solid starting point for subsequent experiments.

The second stage refined the mutation process. It used narrower parameter bounds to optimize mutation rate, the number of mutations per candidate, and related parameters. This step aimed to strike a balance between exploration and exploitation, enhancing the GA’s capacity to discover high-quality solutions without converging prematurely.

All experiments ran on Linux-based high-performance machines in the CIP-Pool at Ludwig Maximilian University, using SLURM for job scheduling. SLURM’s built-in seeding ensured reproducibility, with each experiment executed on seeds one through four in random order.

\section{Results}\label{sec:results}
This work examines how different mutation strategies affect quantum state preparation using a GA. The flexible environment introduced in earlier sections generated extensive data through numerous experiments, which were then evaluated for insights into the effectiveness of each strategy. A detailed description of the experimental setup, including computational resources and dataset generation, can be found in \cref{sec:exp-setup}.

Rather than comparing the GA’s performance to other methods, we focus on how each mutation strategy performs within the same GA framework. We infer effectiveness through well-known metrics, with fidelity as a primary measure of optimization success.

\subsection{Genetic Algorithm Performance}\label{sec:genetic_algorithm_performance}
\begin{figure}[htb]
 \centering
    \includegraphics[page=1,width=\linewidth]{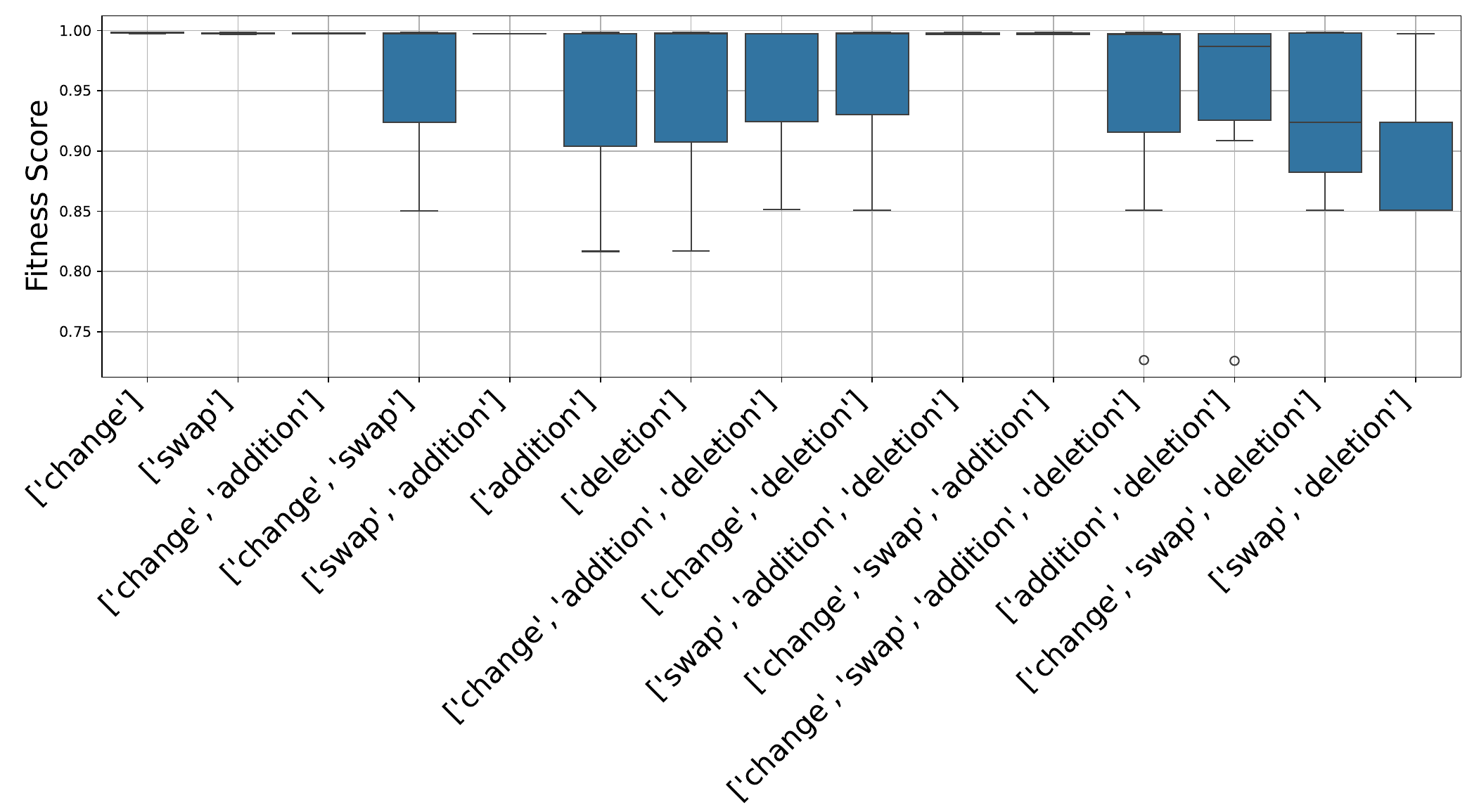}
 \caption{Histogram of average final fitness scores for various mutation strategy combinations on four qubits.}
 \Description{Histogram of average final fitness scores for various mutation strategy combinations on four qubits}
 \label{fig:combine_4w}
\end{figure}

\begin{figure}[htb]
 \centering
    \includegraphics[page=2,width=\linewidth]{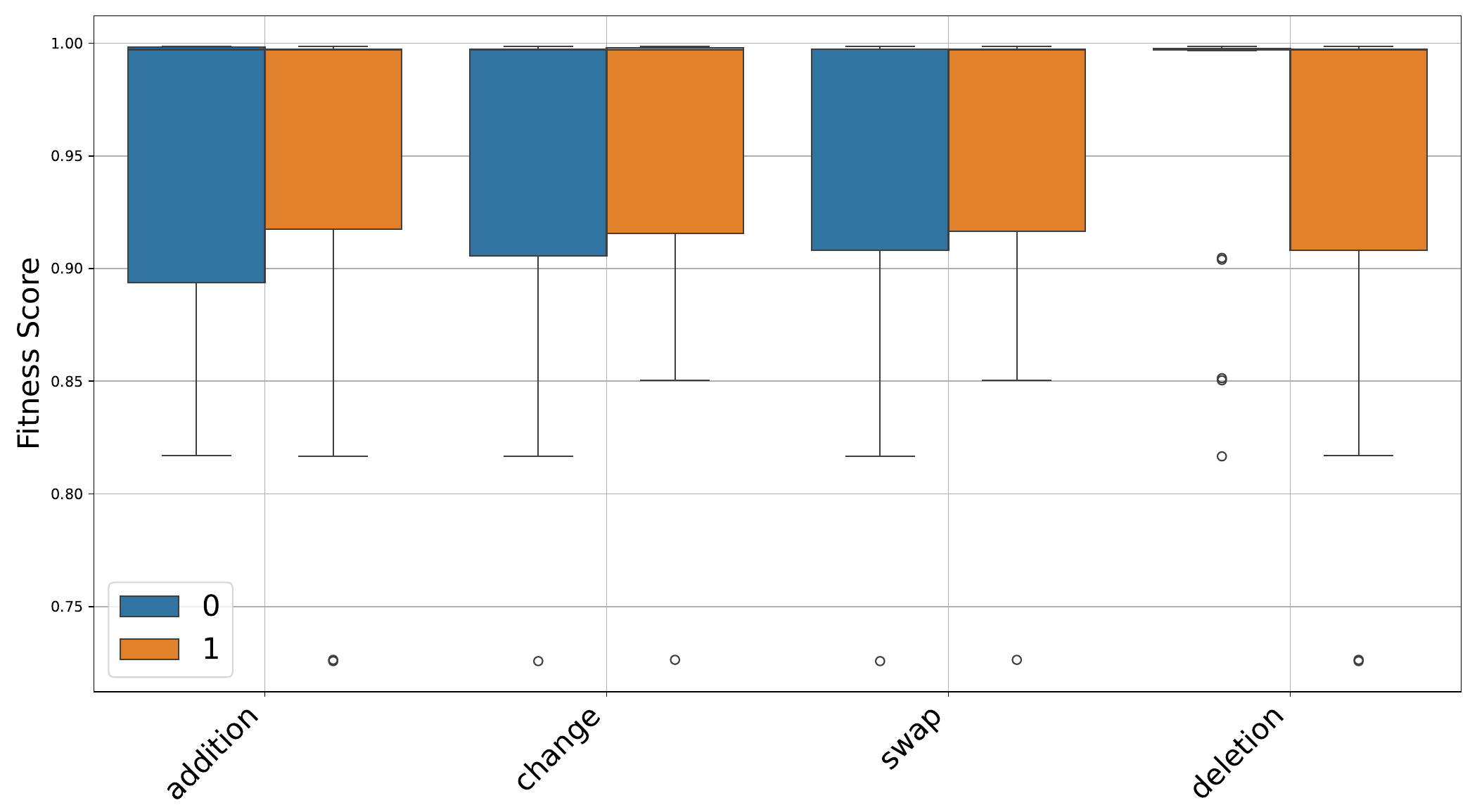}
 \caption{Fitness score comparison on four qubits, showing whether a strategy was present in a trial. Most results approach optimal values.}
 \Description{Fitness score comparison on four qubits, showing whether a strategy was present in a trial}
 \label{fig:single_4w}
\end{figure}

We first tested the GA with target states using different qubit counts to explore how complexity influences algorithmic performance. Four-qubit and six-qubit target states emerged as especially informative. Four-qubit systems, while consistently solved by the GA, did not present significant challenges to distinguish among mutation strategies. \cref{fig:combine_4w} and \cref{fig:single_4w} show that most strategies performed near optimality for four qubits, offering limited insights into their comparative strengths.

Because four qubits proved too simple for a meaningful test, we shifted to six-qubit target states, which provided higher complexity and revealed clearer differences in the performance of various mutation strategies.

\subsection{Impact of Mutation Strategies}\label{sec:impact_of_mutation_strategies}
We evaluated different mutation strategies for six-qubit target states using a dataset of 150 distinct, optimized circuits. Each target state underwent 150 GA generations. This experiment aimed to identify whether specific strategies or strategy combinations yield superior fitness outcomes. We used diverse hyperparameter values to minimize biases in parameter settings. \cref{strategy_combination_table} summarizes the results of 300 trials under different conditions.

\begin{table}[htb]
    \centering
    \caption{Results from 300 trials with different parameter settings chosen by Optuna. The highest mean and 75th percentile appear in the swap, addition combination, while swap, addition, deletion yields the highest median and 25th percentile.}
    \begin{tabularx}{\linewidth}{@{}Xrrrr@{}}
    \toprule
    \textbf{Strategy} & \textbf{Mean} & \textbf{Median} & \textbf{25th Pctl} & \textbf{75th Pctl} \\
    \midrule
    deletion (del) & 0.3432 & 0.2803 & 0.2452 & 0.3488\\
    addition (add) & 0.2939 & 0.2725 & 0.2614 & 0.3435\\
    change (ch) & 0.2660 & 0.2557 & 0.2215 & 0.3156\\
    swap (sw) & 0.3070 & 0.2480 & 0.2302 & 0.2973\\
    ch, add & 0.3209 & 0.2895 & 0.2414 & 0.3168\\
    \textbf{sw, add} & \textbf{0.3598} & 0.2864 & 0.2387 & \textbf{0.4243}\\
    \textbf{sw, del} & 0.3531 & 0.2825 & 0.2495 & 0.3626\\
    ch, del & 0.3298 & 0.2704 & 0.2607 & 0.3088\\
    ch, sw & 0.3281 & 0.2684 & 0.2483 & 0.3096\\
    add, del & 0.2845 & 0.2628 & 0.2587 & 0.2844\\
    \textbf{sw, add, del} & 0.3424 & \textbf{0.2962} & \textbf{0.2630} & 0.3554\\
    ch, sw, add & 0.3207 & 0.2833 & 0.2506 & 0.2980\\
    ch, add, del & 0.2740 & 0.2746 & 0.2416 & 0.2938\\
    ch, sw, del & 0.3289 & 0.2574 & 0.2260 & 0.3674\\
    ch, sw, add, del & 0.2827 & 0.2650 & 0.2420 & 0.2828\\
    \bottomrule
    \end{tabularx}
    
    \label{strategy_combination_table}
\end{table}

\begin{figure}[htb]
 \centering
    \includegraphics[page=5,width=\linewidth]{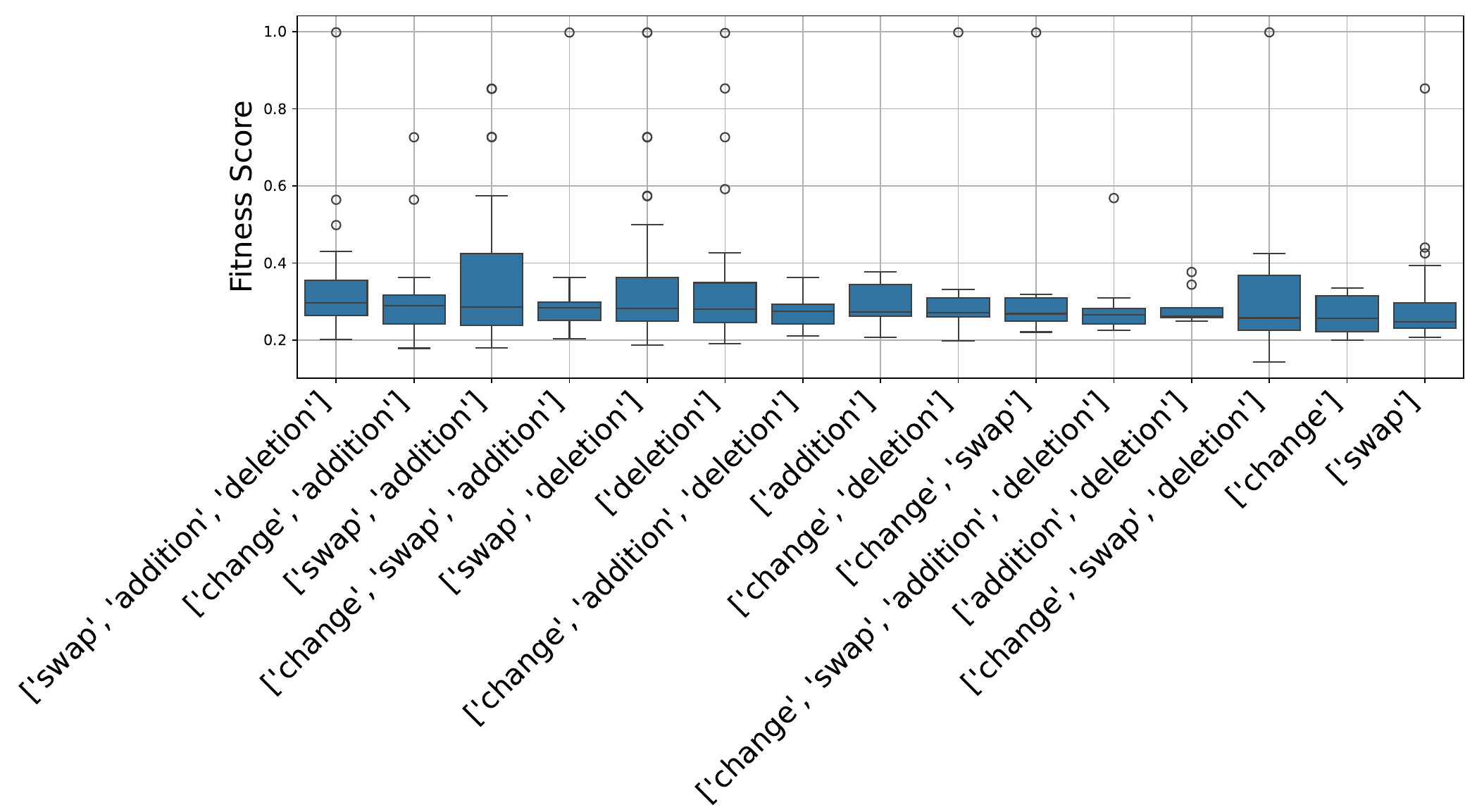}
 \caption{Histogram of average final fitness scores for various strategy combinations on six qubits.}
 \Description{Histogram of average final fitness scores for various strategy combinations on six qubits}
 \label{fig:combine_6w}
\end{figure}

\begin{figure}[htb]
 \centering
    \includegraphics[page=14,width=\linewidth]{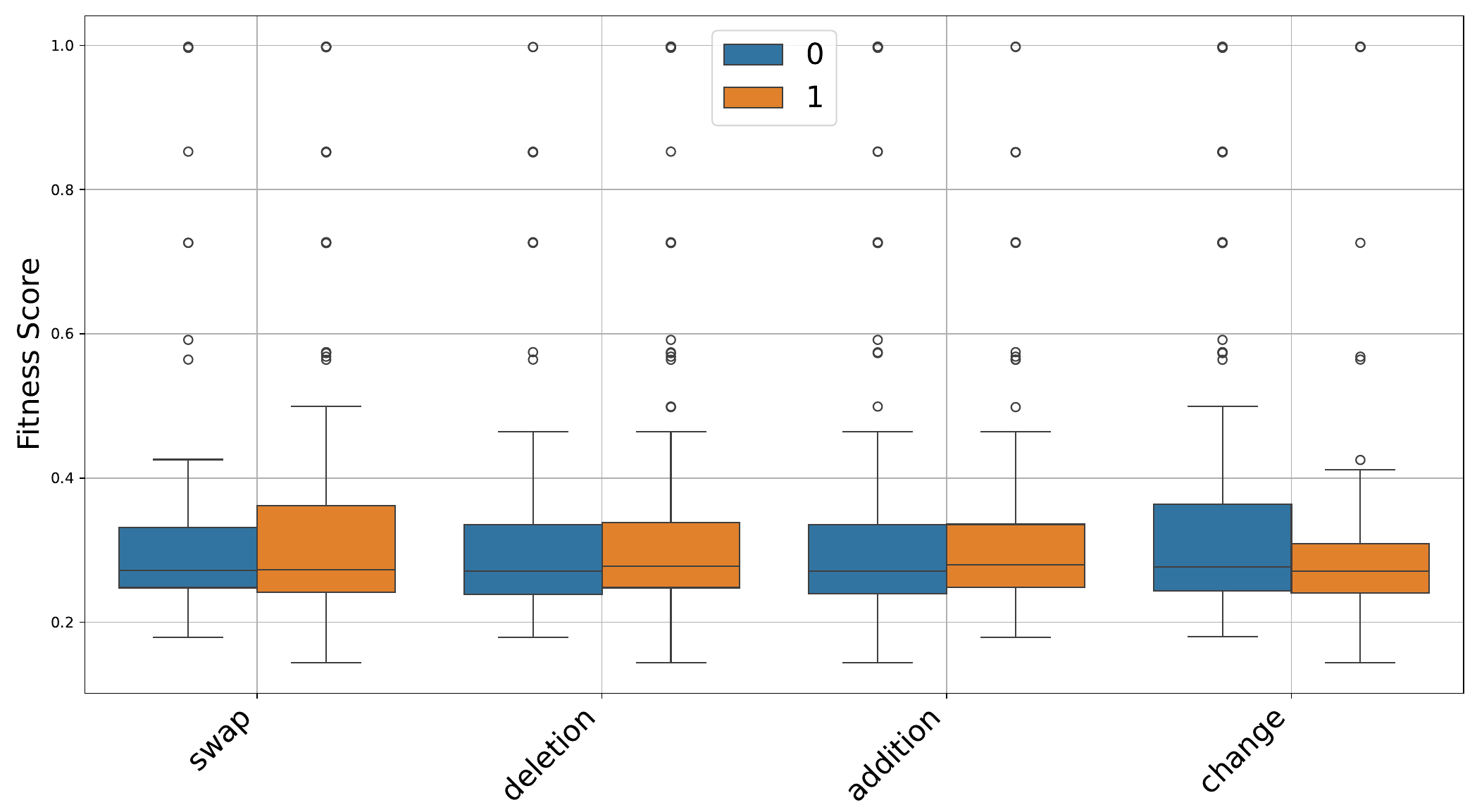}
 \caption{Effect on fitness scores if a strategy was present in a trial for six qubits. Outliers likely reflect the inherent randomness of the GA.}
 \Description{Effect on fitness scores if a strategy was present in a trial for six qubits}
 \label{fig:single_6w}
\end{figure}

\paragraph{Individual Strategies.}
\cref{fig:single_6w} shows that including a mutation strategy typically boosts performance relative to not using it, except for \emph{change}, which lowers overall fitness. \emph{Swap} increases variability the most, while \emph{change} reduces it—at the cost of lower mean fitness. Both \emph{deletion} and \emph{addition} have minimal influence on standard deviation.

\paragraph{Combined Strategies.}
\cref{fig:combine_6w} highlights several promising combinations, especially \emph{swap, addition}, \emph{swap, addition, deletion}, and \emph{swap, deletion}. Notably, these do not include \emph{change}, reinforcing that \emph{change} generally underperforms. The strong impact of \emph{swap} is intuitive, because swapping operations can introduce significant restructuring. The use of \emph{addition} or \emph{deletion} helps maintain or adjust circuit complexity.

\paragraph{Summary.}
\emph{Swap, deletion} emerges as a particularly robust strategy, consistently achieving near-top results and reducing circuit depth over time. \emph{Addition} can be useful but constantly increases depth, making \emph{swap, deletion} an efficient choice for retaining strong performance while limiting circuit growth.

\subsection{Impact of Mutation Rate, Population, and Adaptive Mutation}\label{sec:impact_different_parameter}
We also investigated how hyperparameters—mutation rate, population size, and adaptive mutation—affect mutation strategy performance.

\begin{figure}[htb]
 \centering
    \includegraphics[page=9,width=0.9\linewidth]{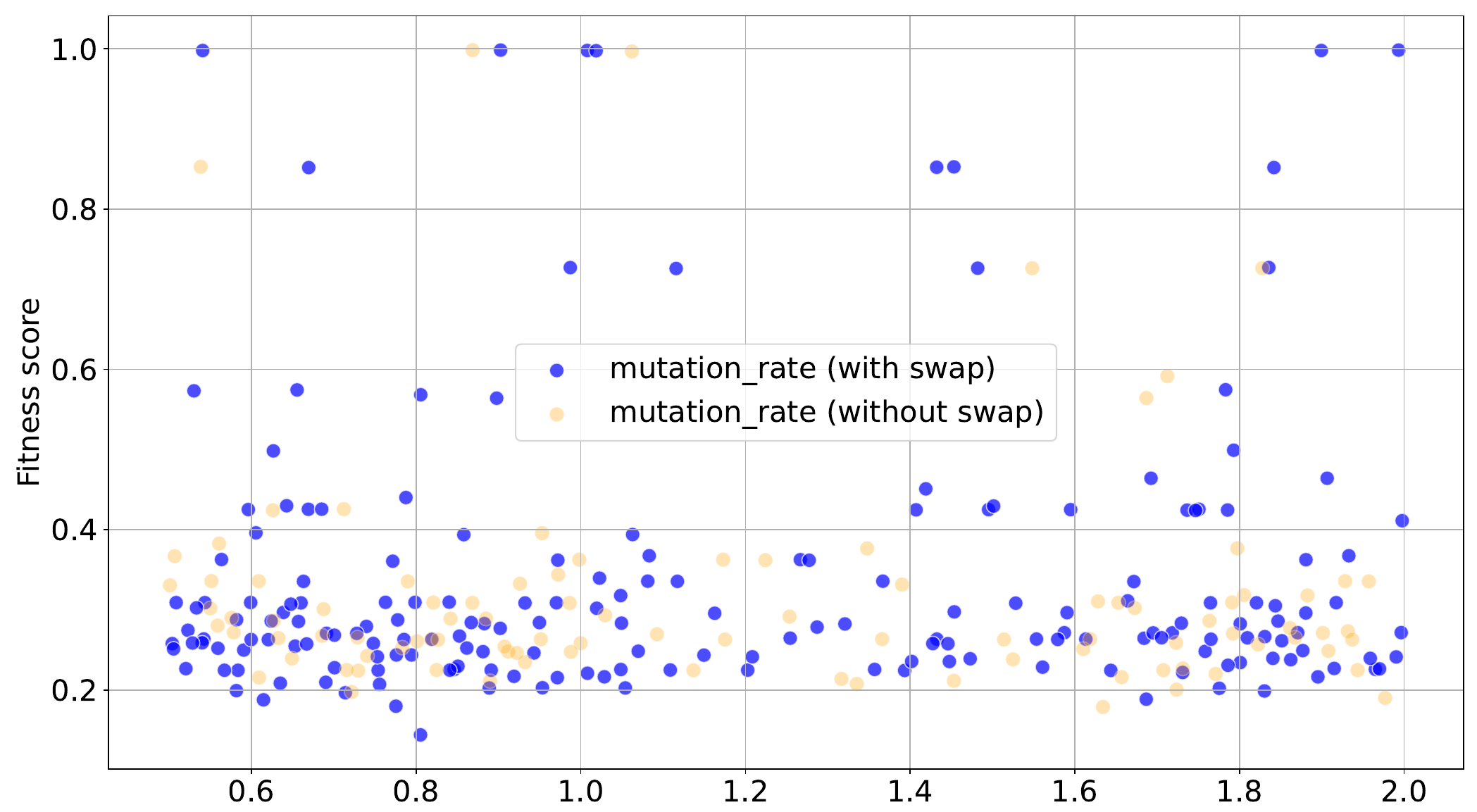}
 \caption{Influence of different mutation rates on \emph{swap}. Rates were already near optimal, resulting in minimal performance changes.}
 \Description{Influence of different mutation rates on swap}
 \label{fig:mu_swap}
\end{figure}

\begin{figure}[htb]
 \centering
    \includegraphics[page=8,width=0.9\linewidth]{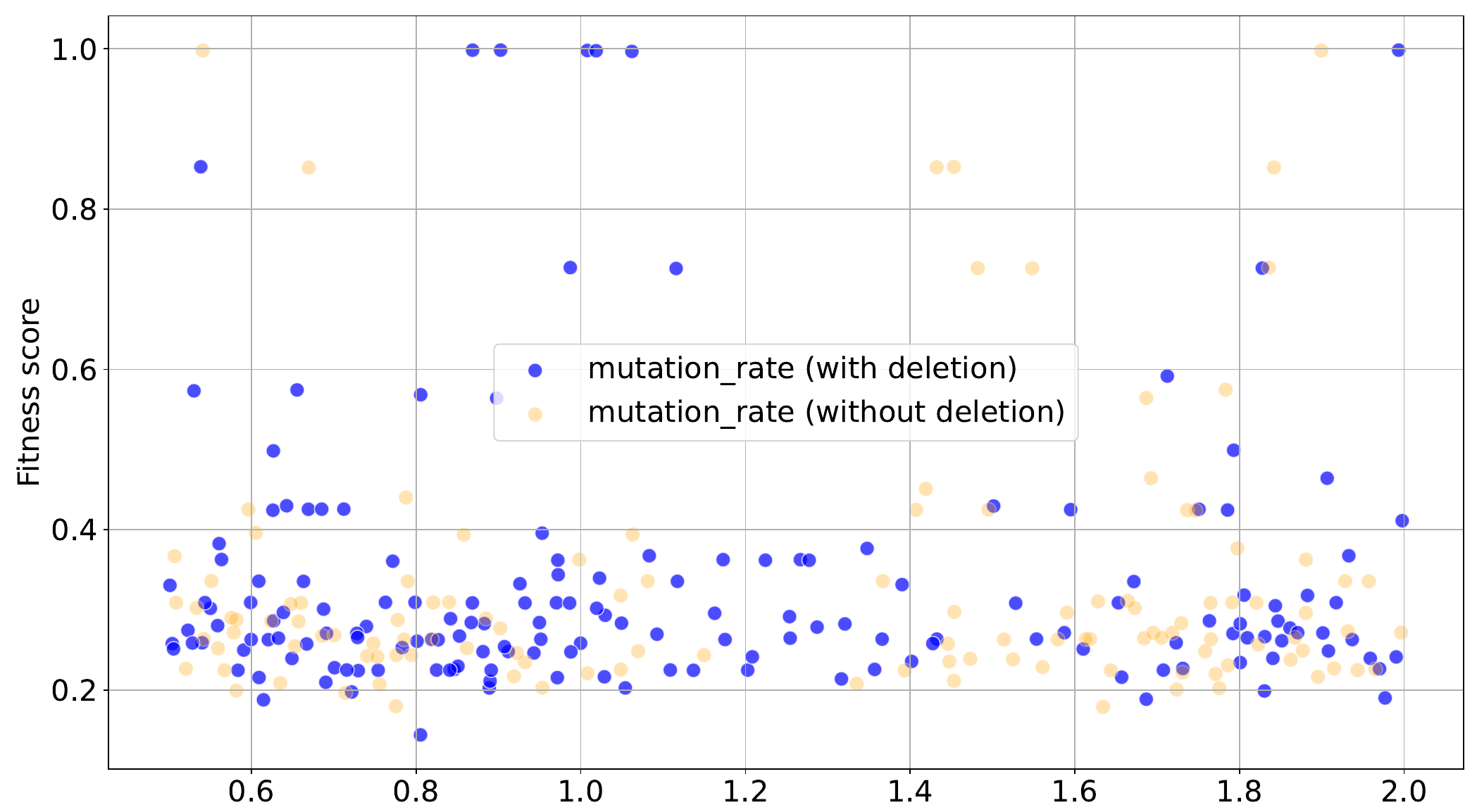}
 \caption{Influence of different mutation rates on \emph{deletion}. Minimal effects are visible.}
 \Description{Influence of different mutation rates on deletion}
 \label{fig:mu_del}
\end{figure}

\paragraph{Mutation Rate.}
\cref{fig:mu_swap} and \cref{fig:mu_del} show that, within the tested ranges, further adjusting mutation rates after narrowing their bounds had little effect on final fitness. This indicates prior tuning already identified rates near local optima.

\begin{figure}[htb]
 \centering
    \includegraphics[page=13,width=0.9\linewidth]{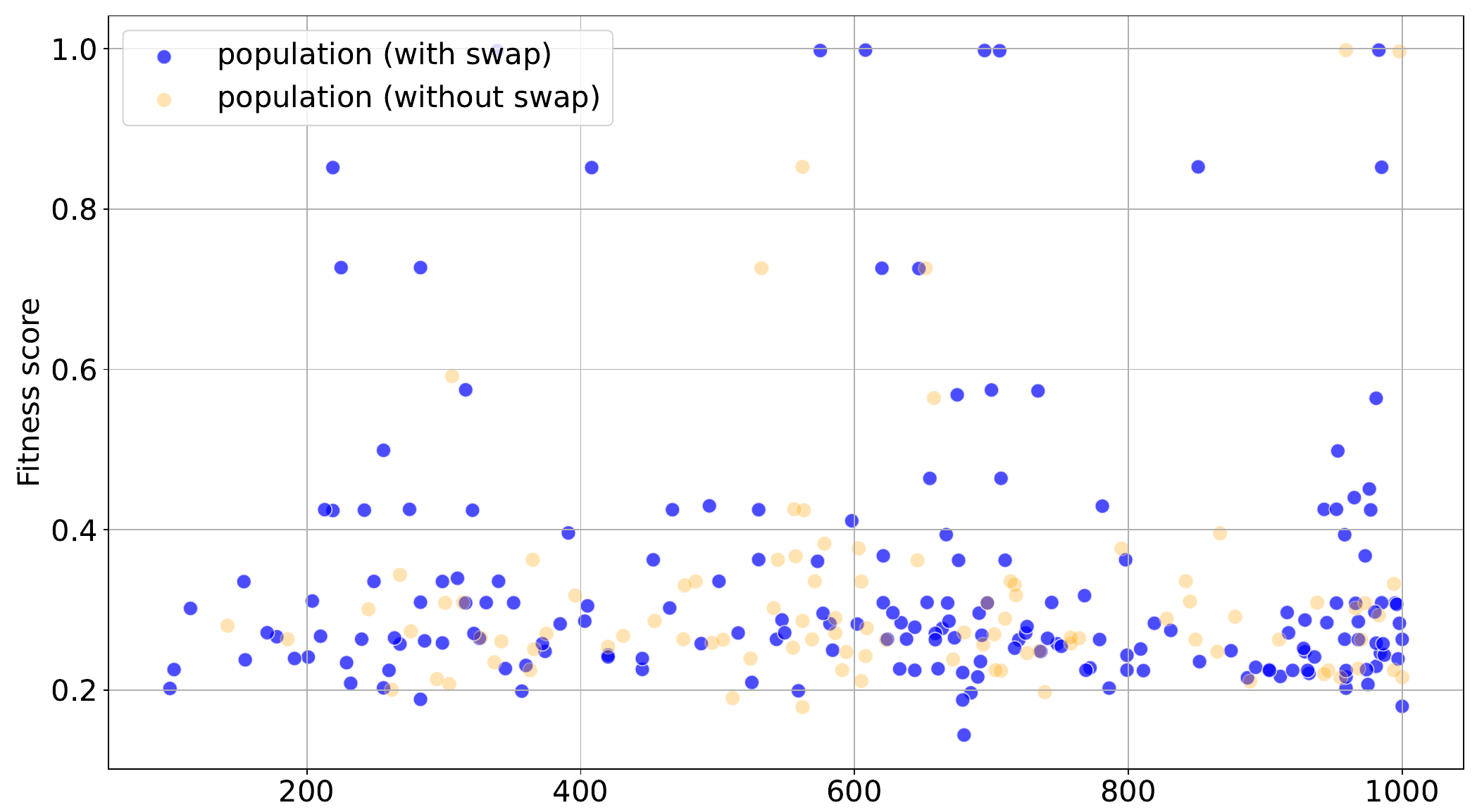}
 \caption{Effect of population size on \emph{swap}. Larger populations slightly increase performance but demand more computational resources.}
 \Description{Effect of population size on swap}
 \label{fig:pop_swap}
\end{figure}

\begin{figure}[htb]
 \centering
    \includegraphics[page=12,width=0.9\linewidth]{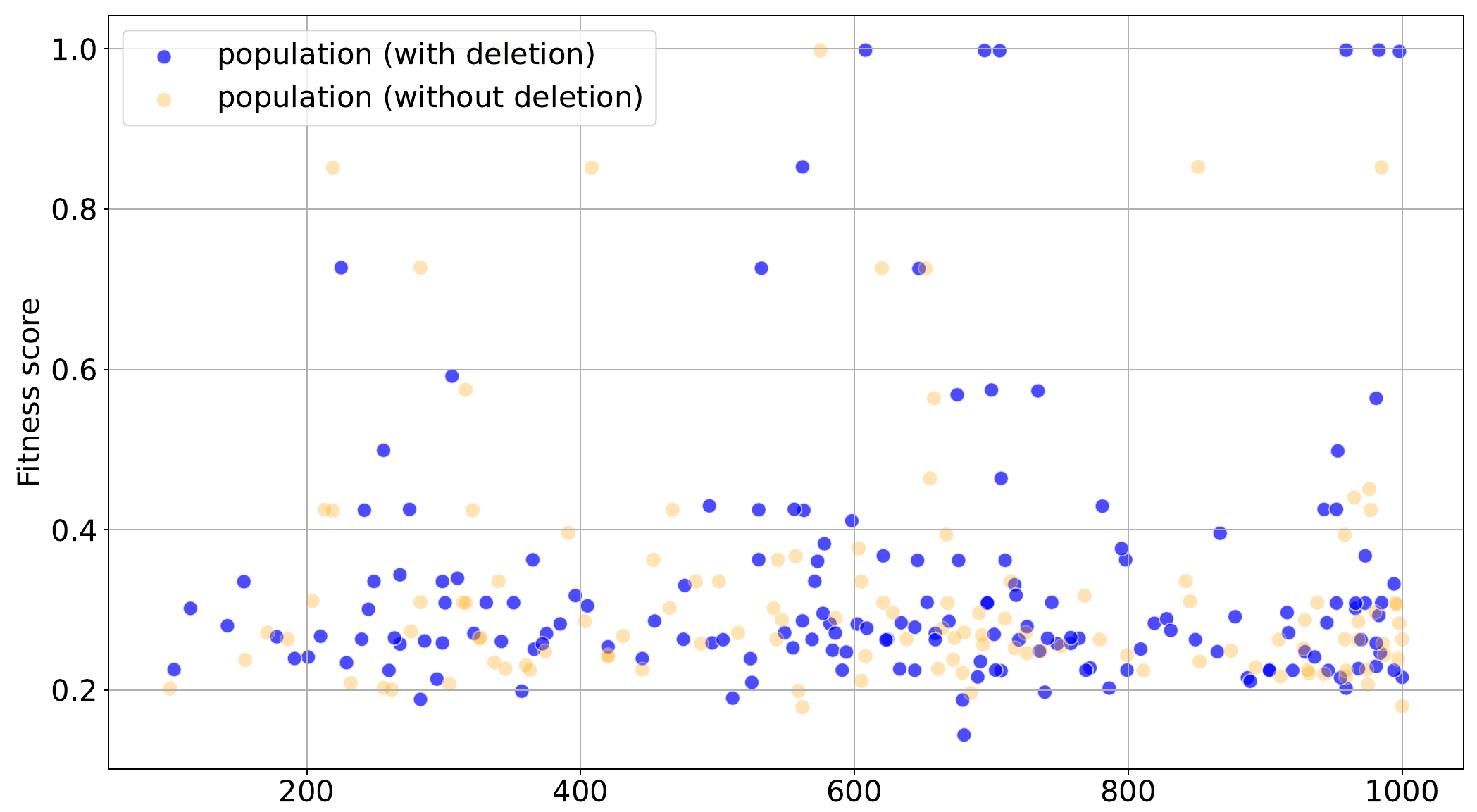}
 \caption{Effect of population size on \emph{deletion}, showing a slight improvement at higher population sizes.}
 \Description{Effect of population size on deletion}
 \label{fig:pop_del}
\end{figure}

\paragraph{Population Size.}
\cref{fig:pop_swap} and \cref{fig:pop_del} show a minor performance boost when increasing the population size, but the gains do not justify the substantial rise in computational costs. Generally, allocating resources to run more generations may be more beneficial than increasing population size.

\begin{figure}[htb]
 \centering
    \includegraphics[page=4,width=0.9\linewidth]{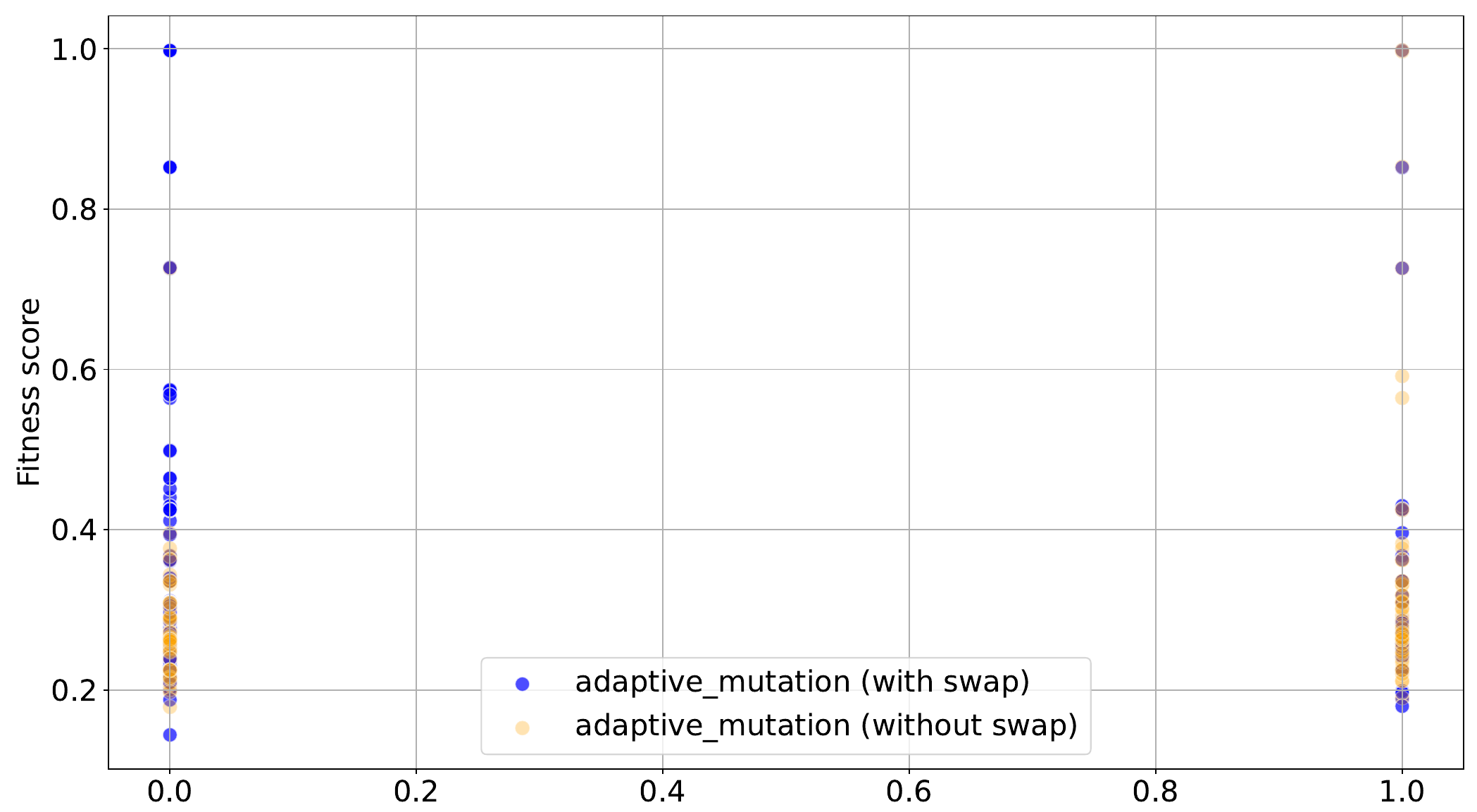}
 \caption{Impact of adaptive mutation on \emph{swap}. Although variability decreases, final fitness also trends lower.}
 \Description{Impact of adaptive mutation on swap}
 \label{fig:ada_swap_6w}
\end{figure}

\begin{figure}[htb]
 \centering
    \includegraphics[page=3,width=0.9\linewidth]{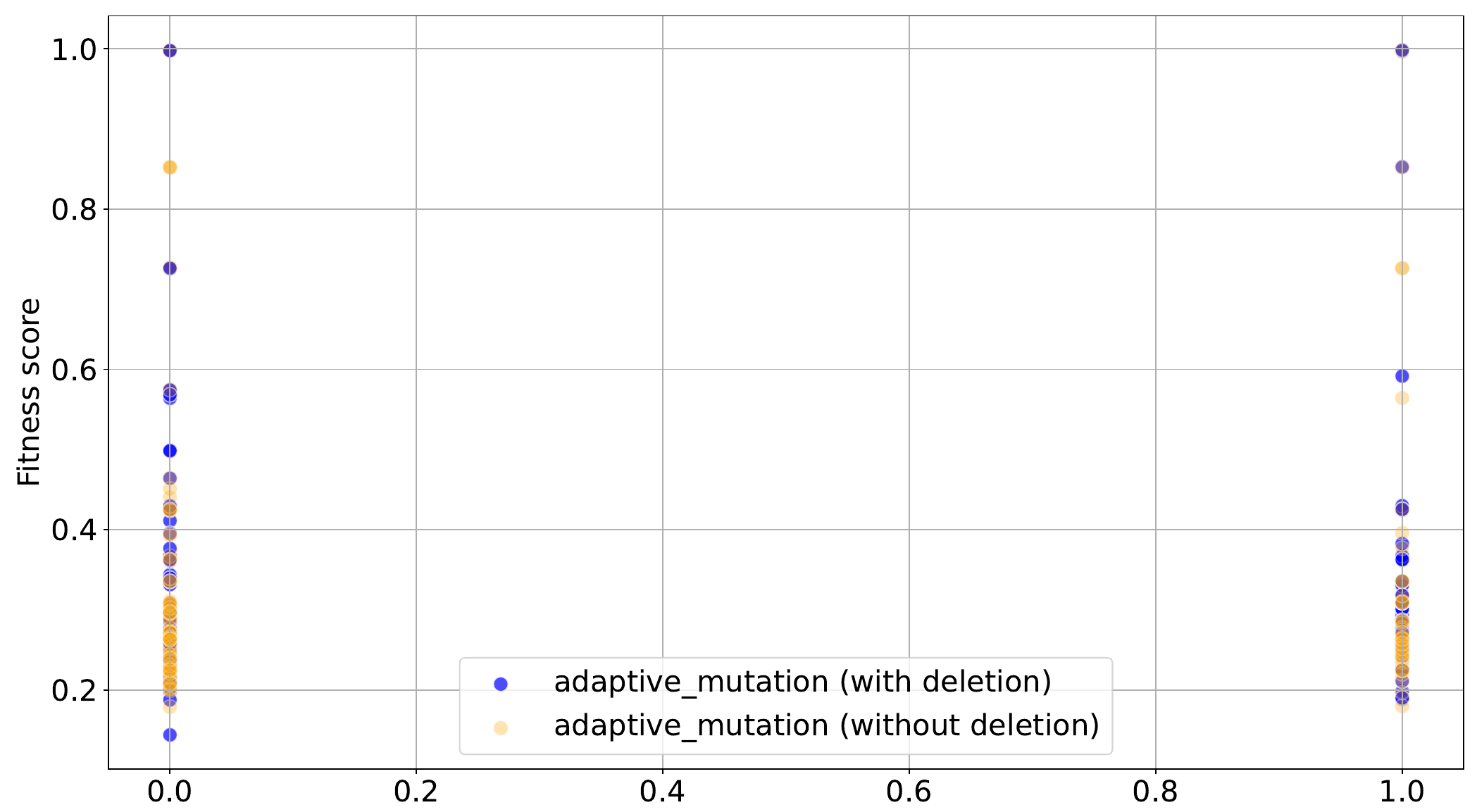}
 \caption{Impact of adaptive mutation on \emph{deletion}. Similar patterns of reduced variability coincide with reduced fitness.}
  \Description{Impact of adaptive mutation on deletion}
 \label{fig:ada_del_6w}
\end{figure}

\paragraph{Adaptive Mutation.}
\cref{fig:ada_swap_6w} and \cref{fig:ada_del_6w} illustrate that adaptive mutation reduces variance but tends to yield lower overall fitness. Although it balances exploration and exploitation by altering rates in response to population diversity and convergence, those adjustments appear to curb performance. A well-tuned static rate often suffices and may simplify implementation while producing stronger final results.

\section{Conclusion}\label{sec:conclusion}

This work investigated how different mutation strategies affect GA performance in quantum state preparation. We conducted extensive experiments in a highly parameterized quantum circuit environment, concluding that the \emph{swap, deletion} strategy generates optimized quantum circuits with the greatest efficiency.

The rising demand for automated circuit synthesis and optimization in NISQ hardware motivated this research. By examining various mutation strategies and their interactions with quantum circuits, we provided empirical data that highlight both the strengths and limitations of GA-based approaches. Our flexible environment and comprehensive dataset represent a main contribution of this work, enabling systematic GA evaluations and offering a resource for further improvements in quantum circuit optimization.

Hardware constraints limited the number of experiments we could conduct. Although we aimed for thorough testing within these constraints, longer trials and larger populations may provide deeper insights. While the findings may not have immediate large-scale effects, identifying potential improvements in the rapidly evolving quantum computing field could lead to significant long-term impact.

Future investigations could explore larger quantum systems, bigger populations, and direct comparisons with alternative optimization methods. Further refining adaptive mutation schemes, perhaps by integrating dynamic heuristics, may also improve convergence speed and performance. We hope that these results encourage ongoing research into evolutionary algorithms for quantum computing, thereby advancing more efficient and fully automated quantum algorithms.

%%
%% The acknowledgments section is defined using the "acks" environment
%% (and NOT an unnumbered section). This ensures the proper
%% identification of the section in the article metadata, and the
%% consistent spelling of the heading.
% \begin{acks}
% To the best projects in the world.
% \end{acks}

%%
%% The next two lines define the bibliography style to be used, and
%% the bibliography file.
\bibliographystyle{ACM-Reference-Format}
\bibliography{main}

%%
%% If your work has an appendix, this is the place to put it.
% \appendix

% \section{Research Methods}

% \subsection{Part One}

% Lorem ipsum dolor sit amet, consectetur adipiscing elit. Morbi
% malesuada, quam in pulvinar varius, metus nunc fermentum urna, id
% sollicitudin purus odio sit amet enim. Aliquam ullamcorper eu ipsum
% vel mollis. Curabitur quis dictum nisl. Phasellus vel semper risus, et
% lacinia dolor. Integer ultricies commodo sem nec semper.

% \subsection{Part Two}

% Etiam commodo feugiat nisl pulvinar pellentesque. Etiam auctor sodales
% ligula, non varius nibh pulvinar semper. Suspendisse nec lectus non
% ipsum convallis congue hendrerit vitae sapien. Donec at laoreet
% eros. Vivamus non purus placerat, scelerisque diam eu, cursus
% ante. Etiam aliquam tortor auctor efficitur mattis.

% \section{Online Resources}

% Nam id fermentum dui. Suspendisse sagittis tortor a nulla mollis, in
% pulvinar ex pretium. Sed interdum orci quis metus euismod, et sagittis
% enim maximus. Vestibulum gravida massa ut felis suscipit
% congue. Quisque mattis elit a risus ultrices commodo venenatis eget
% dui. Etiam sagittis eleifend elementum.

% Nam interdum magna at lectus dignissim, ac dignissim lorem
% rhoncus. Maecenas eu arcu ac neque placerat aliquam. Nunc pulvinar
% massa et mattis lacinia.

\end{document}